# Theoretical study of solvent-mediated Ising-like system: A study for future nanotechnology

Ken-ich Amano[1*]

[1]*Graduate School of Energy Science, Kyoto University, Uji, Kyoto 611-0011, Japan*

*Corresponding author: k-amano@iae.kyoto-u.ac.jp.

**Abstract**

   We theoretically study physical properties of one-dimensionally and regularly placed solutes. The solute is rigid-body, has arrow-like shape, and changes its direction up or down. If the solutes are immersed in continuum solvent, nothing happens in the system. However, the property of the directions differs in granular solvent (e.g., hard-sphere solvent). Depending on distance between the nearest-neighbor solutes, the directional property periodically changes as follows: "ferromagnetic-like" ↔ "random" ↔ "antiferromagnetic-like". Furthermore, the directional property decays into "random" as the distance increases. Studying a newly created nano-system theoretically, it is able to discover a new or interesting property hiding in nano-material world. We believe that such an approach gives a  physics research a new direction and contributes to nanotechnology.

**Main Text**

Recent nanotechnology fabricates various nano-systems [1,2] which are made in accordance with scientific knowledge and experience. However, it is difficult to fabricate a new nano-system by design due to its extreme-smallness and -complexity. For fabrication of the new nano-system, it is important to theoretically comprehend its physical properties beforehand. In addition, it is also important to understand properties of an already-fabricated nano-system. This is because, understanding of the properties enables us to effectively apply the nano-system, for example, in composition of a new nano-machine [3]. In our opinion, solvent-mediated Ising-like (SI) systems will appear in the near future. For example, a system wherein a solvated polymer chain that has regularly arrayed side-molecules is (completely) stretched by optical tweezers [4] is one of the one-dimensional (1D) SI systems as long as the number of orientations for the side-molecule is two and the orientation is changed by rotating around long-axis of the stretched polymer chain (see Fig. 1). Likewise, two-dimensional (2D) and three-dimensional (3D) SI systems can be imagined. To understand properties of SI systems and to discover new or interesting properties in SI systems, we start to study SI systems.

In this letter, we theoretically study directional properties of two *primitive* 1D-SI systems (see Figs. 2(a) and 2(b)). The solutes are rigid-bodies and immersed in hard-sphere solvent in this study. The all-rigid-body system is chosen as the first step study (non-rigid-body system can also be treated). Diameter of the solvent particle ($d_S$) is set at 0.28 nm which is the same as that of a water molecule. Solvent bulk density is expressed as $\rho_S$, and $\rho_S d_S^3$ (dimensionless value) is set at 0.7317 which is the same as that of water under normal condition (1 atm, 298 K). We use two types of the solutes, both of which have arrow-like shapes. The solutes shown in Figs. 2(c) and 2(d) are referred to as conical and cylindrical solutes, respectively. The solutes change their own directions up (↑) or down (↓). The solutes are one-dimensionally and regularly placed in the solvent (centers of the solutes, which are represented as red points, are on a straight line). The number of the solutes $N$ is set at sufficiently large value 100 which is decided by taking account of the stretched polymer chain in some degree. Distance between the nearest-neighbor solutes (red points) is written as $d_2$, and tested range of $d_2/d_S$ is 4.0-8.0.

Theoretical treatment of the 1D-SI system is explained as follows. Firstly, we consider partition function of the 1D-SI system. The partition function ($Z$) is written as,

$$Z = \sum_{\sigma_1=\pm 1} \cdots \sum_{\sigma_N=\pm 1} \int \cdots \int \exp\{-E(\mathbf{r}_1 \cdots \mathbf{r}_V, \sigma_1 \cdots \sigma_N)/(k_B T)\} d\mathbf{r}_1 \cdots d\mathbf{r}_V, \quad (1)$$

where $\sigma_i$ ($i=1$-$N$) represents direction of the solute at position $i$. When $\sigma$ is $+1$ the direction is ↑, while when $\sigma$ is $-1$ the direction is ↓. $E$ is the instant value of potential energy for a system configuration, $\mathbf{r}_j$ ($j=1$-$V$) 3D coordinates of the $j$-th solvent particle, $k_B$ the Boltzmann constant, and $T$ the absolute temperature. $E$ can be decomposed into configurational energy of the solutes ($E_C$) and the solvent energy ($E_S$),

$$E = E_C(\sigma_1 \cdots \sigma_N) + E_S(\mathbf{r}_1 \cdots \mathbf{r}_V; \sigma_1 \cdots \sigma_N). \quad (2)$$

Hence, the partition function is rewritten as

$$Z = \sum_{\sigma_1=\pm 1} \cdots \sum_{\sigma_N=\pm 1} \exp\{-E_C(\sigma_1 \cdots \sigma_N)/(k_B T)\} \int \cdots \int \exp\{-E_S(\mathbf{r}_1 \cdots \mathbf{r}_V; \sigma_1 \cdots \sigma_N)/(k_B T)\} d\mathbf{r}_1 \cdots d\mathbf{r}_V. \quad (3)$$

Here, the integrations part in Eq. (3) is a partition function for the solvent, $Z_S$. Since $\mu_S/(k_B T) = -\ln Z_S$ ($\mu_S$ is the solvent free energy of a solutes configuration) and $E_C$ is zero in the all-rigid-body system, Eq. (3) is rewritten as [5-7]

$$Z = \sum_{\sigma_1=\pm 1} \cdots \sum_{\sigma_N=\pm 1} \exp\{-\mu_S(\sigma_1 \cdots \sigma_N)/(k_B T)\}. \quad (4)$$

The solvent free energy consists of the free energy of the pure solvent ($\mu_{PS}$) and the solvation free energy of a solutes configuration ($\mu_{SOL}$). That is, $\mu_S = \mu_{PS} + \mu_{SOL}$. When the solutes are sufficiently separated, $\mu_{SOL}$ is given by $N\mu_1$ ($\mu_1$ is the solvation free energy of the solute). Since $\mu_1$ is the solvation free energy of the *one* solute with computationally available size, it can be calculated by 3D integral equation theory [8-14], a statistical-mechanical theory for fluids (see Supplemental Material [15]). However, the distance ($d_2/d_S$) we consider is from 4.0 to 8.0, and so there should be unignorable correlations among the solutes ($\mu_{SOL}$ cannot be given by $N\mu_1$). To overcome this problem, an approximation that only interactions between the nearest-neighbor solutes are effective is introduced. That is, interactions between the $i$-th and $(i+n)$-th ($n \geqq 2$) solutes are ignored. (Validity of this approximation is discussed in Supplemental Material [15].) Thus, $\mu_S$ can be given by

$$\mu_S = \mu_{PS} + N\frac{\mu_{2P} + \mu_{2A}}{2} + \frac{\mu_{2P} - \mu_{2A}}{2}\sum_{i=1}^{N}\sigma_i\sigma_{i+1} \qquad (\sigma_{i+1} = \sigma_1), \qquad (5)$$

where, $\mu_{2P}$ and $\mu_{2A}$ are the solvation free energies of the two solutes at parallel (↑↑) and antiparallel (↑↓) orientations, respectively, and ($\sigma_{N+1}=\sigma_1$) is the boundary condition. Solvation free energy of the two separated solutes ($\mu_{2P}$ or $\mu_{2A}$) can be calculated using 3D integral equation theory [8-14] by regarding the separated solutes as the *one* solute [10] as long as the solutes are not extremely large (i.e., the *one* solute is computationally available size). Substituting Eq. (5) into Eq. (4), the partition function is rewritten as

$$Z = \exp\left\{-(\mu_{PS} + N\frac{\mu_{2P} + \mu_{2A}}{2})/(k_B T)\right\} \sum_{\sigma_1=\pm 1}\cdots\sum_{\sigma_N=\pm 1} \exp\left\{-\frac{\Delta\mu_{PA}}{2}\sum_{i=1}^{N}\sigma_i\sigma_{i+1}/(k_B T)\right\}, \qquad (6)$$

where $\Delta\mu_{PA}$ is $\mu_{2P}-\mu_{2A}$, which is the solvation free energy difference between parallel and antiparallel orientations. The summations part in Eq. (6) corresponds to a partition function for 1D Ising model [16-18]. The partition function is known to be solved exactly, and then Eq. (6) is rewritten as [16-18]

$$Z = \exp\left\{-\left(\mu_{PS} + N\frac{\mu_{2P} + \mu_{2S}}{2}\right)/(k_B T)\right\} \cdot \left[\left\{2\cosh\left(-\frac{\Delta\mu_{PA}}{2k_B T}\right)\right\}^N + \left\{2\sinh\left(-\frac{\Delta\mu_{PA}}{2k_B T}\right)\right\}^N\right], \qquad (7)$$

To investigate a physical property of the 1D-SI system, we calculate directional correlation between the nearest-neighbor solutes. (In a general study of Ising model, directional property of magnetic spins is discussed with magnetization intensity or magnetic susceptibility. However, a measure that corresponds to magnetization intensity or magnetic susceptibility cannot be defined in the all-rigid-body system. Therefore, we discuss the directional correlation between the nearest-neighbor solutes.) Here, the directional correlation <$\sigma_i\sigma_{i+1}$> is obtained as follows:

$$\langle\sigma_i\sigma_{i+1}\rangle = \frac{1}{N}\cdot\frac{\partial \ln Z}{\partial L} = \frac{\cosh^{N-2}(L)+\sinh^{N-2}(L)}{\cosh^{N}(L)+\sinh^{N}(L)}\cdot\cosh(L)\sinh(L). \tag{8}$$

where "<$X$>" is ensemble average of $X$ and $L=-\Delta\mu_{PA}/(2k_BT)$. When $L$ is equal to $+\infty$, <$\sigma_i\sigma_{i+1}$> becomes $+1$, and the directional correlation is completely parallel. In contrast, when $L$ is equal to $-\infty$, <$\sigma_i\sigma_{i+1}$> becomes $-1$. The directional correlation is completely antiparallel. However, the system considered here is not such an extreme condition, because $L$ is finite value. Therefore, the value of the directional correlation never be equal to $\pm 1$, which is larger than $-1$ and smaller than $+1$.

Here, we show the results. In Fig. 3(a), the directional correlations of the conical and cylindrical solutes are shown with triangle and circle points, respectively. As observed in Fig. 3(a), the two curves oscillate around zero, and their pitches are about $d_S$. The directional correlations are periodically changed as follows: "parallel-dominant" ↔ "random" ↔ "antiparallel-dominant". (When the value of <$\sigma_i\sigma_{i+1}$> is within $\pm 0.5$, the term "dominant" should not be used, but we use it for simple naming.) In addition, their amplitudes decay into zero as $d_2$ increases (i.e., the directional correlation vanishes). These behaviors derive from oscillatory fluctuation of $\Delta\mu_{PA}$ (=$\mu_{2P}-\mu_{2A}$), and it is similar to a distribution function of a solvent [8,9,19,20] and solvent-mediated potential of mean force between two arbitrary solutes [7-14,21]. The behaviors of the directional correlation are interesting, because the system considered here is the all-rigid-body system with no magnetic or electrostatic interaction. Although the results are not shown here, we have found the same behaviors in the similar 1D-SI systems. The behaviors are omnipresent at least in the all-rigid-body 1D-SI systems. In our opinion, the behaviors exist also in some realistic systems.

The results above are calculated in the solvent with higher density ($\rho_S d_S^3$=0.7317). In this paragraph, we briefly show an effect of the solvent density by varying it ($\rho_S d_S^3$=0.0, 0.1, 0.2, 0.3, 0.4, 0.5, 0.6, 0.7, and 0.8). In Fig. 3(b), the directional correlations of the conical and cylindrical solutes are shown with triangle and circle points, respectively, where $d_2$ is set at $4.0d_S$. As observed in Fig. 3(b), the value of <$\sigma_i\sigma_{i+1}$> is changed from positive to negative and it converges into zero as the solvent density decreases. When the solvent density is higher, parallel-dominant condition appears, while when the solvent density is lower, antiparallel-dominant condition weakly appears.

Mechanism of the signal change from positive to negative in Fig. 3(b) is explained by using the cylindrical solutes (we remark that $d_2$ is $4.0d_S$ here). When the solvent density is higher, some of the solvent particles are effectively packed between the solutes with parallel orientation (see Fig. 4(a)). The effective packing enables the other solvent particles to displace themselves in larger space, leading to gain in translational entropy of the solvent. On the other hand, the solvent particles cannot effectively be packed between the solutes with antiparallel orientation (see Fig. 4(b)). There are spaces (depicted by diagonal lines) that are unavailable for translational displacement of the solvent particles. It leads to loss in translational entropy of the solvent. In this case, $\mu_{2P}$ is smaller than $\mu_{2A}$ (i.e., $\Delta\mu_{PA}<0$ and $L>0$), and hence the value of $<\sigma_i\sigma_{i+1}>$ becomes positive. Also when the solvent density is lower, the packing effect exists, but the effect is slightly changed. Here, we consider solvent-mediated potential of mean force ($\Phi$) between two spherical solutes with $6d_S$ in diameters to simplify the present explanation (see Figs. 4(c) and 4(d)). Figure 4(d) shows $\Phi/(k_BT)$ vs interval between the closest surfaces of the spherical solutes ($I_2$). Solid, broken, and dashed curves are that calculated within $\rho_S d_S^3 =0.3$, 0.5, and 0.7, respectively, and they are calculated by using the radial symmetric integral equation theory for spherical particles [22]. As observed in Fig. 4(d), positions of the local minima near $I_2/d_2=1.0$ (marked in square) is moved away from $I_2/d_2=1.0$ as the solvent density decreases. When $\rho_S d_S^3 =0.7$ and $I_2/d_2=1.0$, the two spherical solutes is stabilized, because the value of $\Phi/(k_BT)$ is negative. On the other hand, when $\rho_S d_S^3 =0.3$ and $I_2/d_2=1.0$, they are destabilized due to the positive value of $\Phi/(k_BT)$. To stabilize them, $I_2$ should be increased little bit. This tendency appears also in the case of the cylindrical solutes. That is, the solutes with parallel orientation ($d_2=4.0d_S$) are no longer stabilized in the lower density solvents. The solutes with antiparallel orientation become more stable than that with parallel orientation in this case. Then, $\mu_{2P}$ becomes larger than $\mu_{2A}$ (i.e., $\Delta\mu_{PA}>0$ and $L<0$), and the value of $<\sigma_i\sigma_{i+1}>$ becomes negative. These are mechanisms of the signal change from positive to negative in Fig. 3(b).

In summary, we have theoretically studied solvent-mediated Ising-like (SI) system. The conical and cylindrical solutes are one-dimensionally and regularly placed in the hard-sphere solvent. It has been found that depending on $d_2$, the directional correlation is periodically changed as follows: "parallel-dominant" ↔ "random" ↔ "antiparallel-dominant". Furthermore, the directional correlation decays into "random" as $d_2$ increases and the solvent density decreases. Studying a newly created nano-system theoretically, we have discovered new or interesting properties hiding in nano-material world. As a next study, we will tackle a system shown in Fig. 1. Effects of temperature and external field on a 1D-SI system will also be studied in the near future. 2D- and 3D-SI systems will also be studied. 2D-SI system is, for instance, related to a system wherein simple solutes (or side-molecules) are two-dimensionally and regularly placed on a sheet of DNA origami [23,24]. Although we do not explain in detail here, our theoretical treatment can be readily extended to a non-rigid-body system. For examples, Lennard-Jones fluid can be used as the solvent by re-employing 3D integral equation theory [7-14], and water (aqueous solution) can be introduced by employing 3D-reference interaction site model [25] or theory of energy representation [26-28]. Therefore, effects of solvophobicity (hydrophobicity) and solvophilicity (hydrophilicity) on a SI system can be studied. A spin with three, four, five, or several orientations (potts spin [29]) is also in our interest. In our opinion, the study of SI systems leads to developments of physics and nanotechnology. We believe that properties which have not been discovered yet still exist in nano-material world. If the new property is theoretically discovered, nanotechnology will be developed more, and the development of the nanotechnology activates physics researches. Thus, our new approach is very important for developments of theoretical and experimental studies.

We appreciate discussions with Hiraku Oshima (Kyoto University). We thank Masahiro Kinoshita (Kyoto University) for providing the basic parts of the computer programs for integral equation theories. This work was supported by Grant-in-Aid for JSPS (Japan Society for the Promotion of Science) fellows.

**References**
[1] T. Liedl, B. Högberg, J. Tytell, D. E. Ingber, and W. M. Shih, Nature Nanotech. **5**, 520 (2010).
[2] P. Guo, Nature Nanotech. **5**, 833 (2010).
[3] K. Amano, D. Miyazaki, L. Fong Fong, P. Hilscher, and T. Sonobe, Phys. Lett. A **375**, 165 (2010).
[4] A. N. Gupta *et al.*, Nature Phys. **7**, 631 (2011).
[5] A. Mitsutake, M. Kinoshita, Y. Okamoto, and F. Hirata, J. Phys. Chem. B **108**,

19002 (2004).

[6] A. Mitsutake and Y. Okamoto, J. Chem. Phys. **130**, 214105 (2009).

[7] K. Amano, H. Oshima, and M. Kinoshita, Chem. Phys. Lett. **504**, 7 (2011).

[8] M. Kinoshita and T. Oguni, Chem. Phys. Lett. **351**, 79 (2002).

[9] M. Kinoshita, J. Chem. Phys. **116**, 3493 (2002).

[10] M. Kinoshita, Chem. Phys. Lett. **387**, 47 (2004).

[11] M. Kinoshita, Chem. Eng. Sci. **61**, 2150 (2006).

[12] K. Amano, T. Yoshidome, M. Iwaki, M. Suzuki, and M. Kinoshita, J. Chem. Phys. **133**, 045103 (2010).

[13] K. Amano and M. Kinoshita, Chem. Phys. Lett. **488**, 1 (2010).

[14] K. Amano and M. Kinoshita, Chem. Phys. Lett. **504**, 221 (2011).

[15] See Supplemental Material at [URL] for details of the calculation (3D integral equation theory) and the approximation (neglect of interactions between the $i$-th and $(i+n)$-th ($n \geqq 2$) solutes).

[16] H. A. Kramers and G. H. Wannier, Phys. Rev. **60**, 252 (1941).

[17] M. Suzuki, B. Tsujiyama, and S. Katsura, J. Math. Phys. **8**, 124 (1967).

[18] R. J. Baxter, *Exactly solved models in statistical mechanics* (Harcout Brace Jovanovich, 1982).

[19] J. L. Yarnell, M. J. Katz, R. G. Wenzel, and S. H. Koenig, Phys. Rev. A. **7**, 2130 (1973).

[20] G. Hura *et al.*, Phys. Chem. Chem. Phys. **5**, 1981 (2003).

[21] J. C. Crocker, J. A. Matteo, A. D. Dinsmore, and A. G. Yodh, Phys. Rev. Lett. **82**, 4352 (1999).

[22] M. Kinoshita, S. Iba, K. Kuwamoto, and M. Harada, J. Chem. Phys. **105**, 7177 (1996).

[23] J. D. Le *et al.*, Nano Lett. **4**, 2343 (2004).

[24] A. M. Hung *et al.*, Nature Nanotech. **5**, 121 (2010).

[25] A. Kovalenco and F. Hirata, Chem. Phys. Lett. **290**, 237 (1998).

[26] N. Matubayasi and M. Nakahara, J. Chem. Phys. **113**, 6070 (2000).

[27] N. Matubayasi and M. Nakahara, J. Chem. Phys. **117**, 3605 (2002).

[28] N. Matubayasi and M. Nakahara, J. Chem. Phys. **119**, 9686 (2003).

[29] F. Y. Wu, Rev. Mod. Phys. **54**, 235 (1982).

**Figure Captions**

FIG. 1. Examples of 1D-SI systems. (a), a polymer chain is (completely) stretched by optical tweezers in solvent. The stretched polymer chain has regularly arrayed side-molecules (ellipses colored yellow-green), the number of orientations for the side-molecule is two. Distance between the nearest-neighbor side-molecules is varied by changing length of the red molecule, as shown in (b). (c), examples of the two orientations of the side-molecule, which are observed from the direction of long-axis of the stretched polymer chain. "u" and "d" mean up and down, respectively. A set of vertical "v" and horizontal "h" orientations are also probable as the two orientations.

FIG. 2. (a) and (b), tested 1D-SI systems where the conical and cylindrical solutes are on-dimensionally and regularly placed in the hard-sphere solvent, respectively. Geometrical details of the conical and cylindrical solutes are drawn in (c) and (d), respectively. Distance between the nearest-neighbor solutes ($d_2$) is that between red points.

FIG. 3. Directional correlation between the nearest-neighbor solutes. $<\sigma_i\sigma_{i+1}>$ vs $d_2/d_S$ is shown in (a), where $\rho_S d_S^3$ is set at 0.7317. $<\sigma_i\sigma_{i+1}>$ vs $\rho_S d_S^3$ is shown in (b), where $d_2/d_S$ is set at 4.0. For judgments of positive and negative values, broken lines ($<\sigma_i\sigma_{i+1}>=0.0$) are inserted in (a) and (b). The results for the conical and cylindrical solutes are shown with triangle and circle points, respectively.

FIG. 4. (a) and (b), packings of the solvent particles between the cylindrical solutes with parallel and antiparallel orientations, respectively. The spaces depicted by diagonal lines are unavailable spaces for translational displacement of the solvent particles. (c), two spherical solutes with $6d_S$ in diameters are immersed in the hard-sphere solvent. $I_2$ is interval between the closest surfaces of the spherical solutes. (d), solvent-mediated potential of mean force between the spherical solutes ($\Phi$) divided by $k_B T$ vs $I_2/d_S$. Solid, broken, and dashed curves are that calculated within $\rho_S d_S^3$=0.3, 0.5, and 0.7, respectively. Square points represent that of the local minima around $I_2/d_S=1.0$. A horizontal broken line ($\Phi/(k_B T)=0.0$) is also inserted.

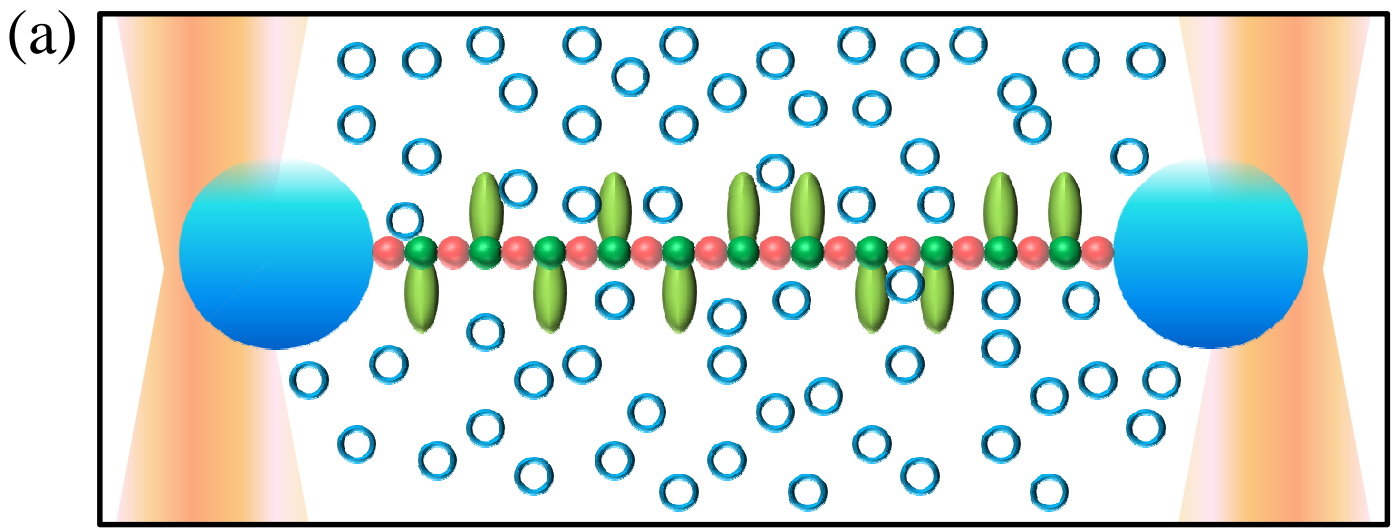

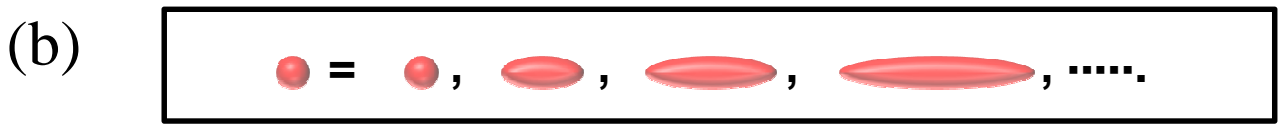

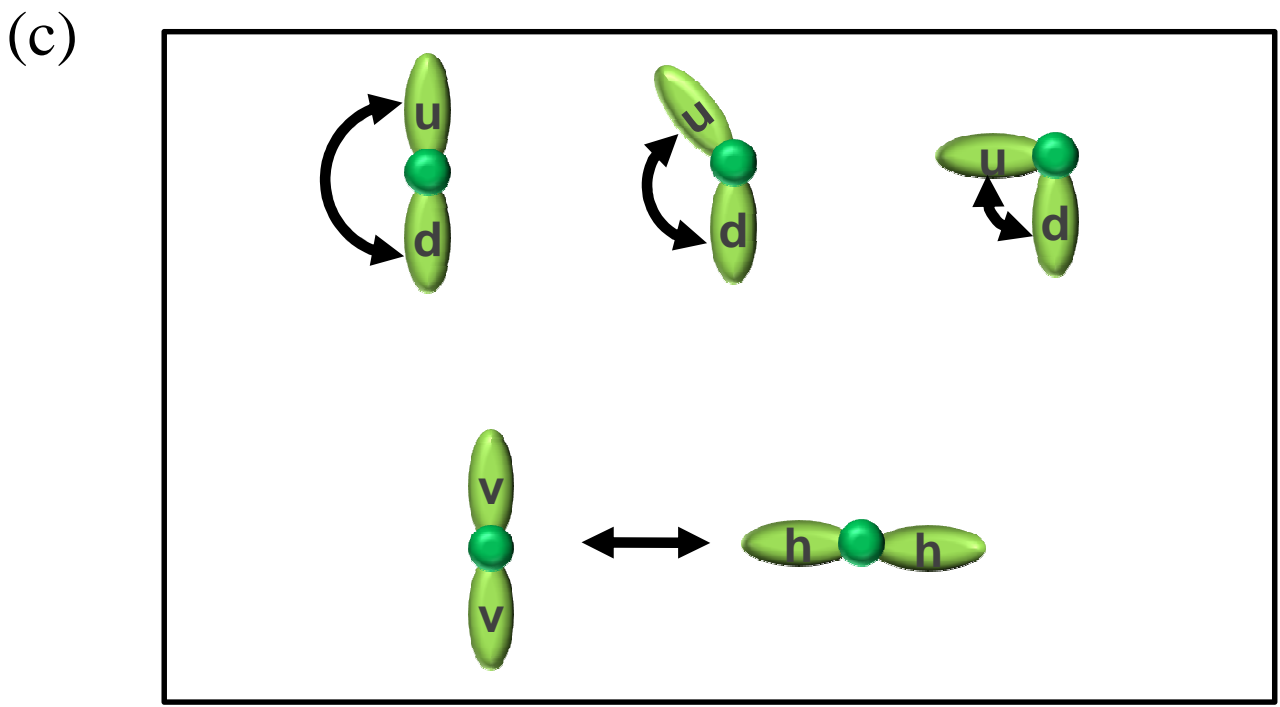

FIG. 1

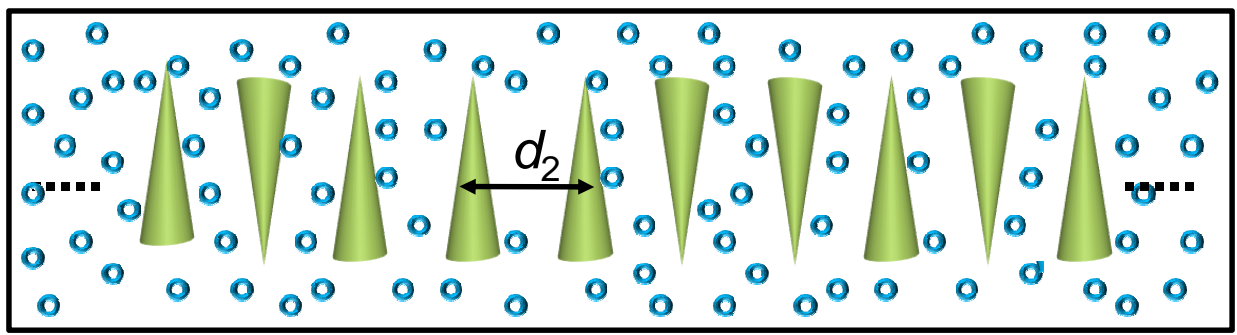

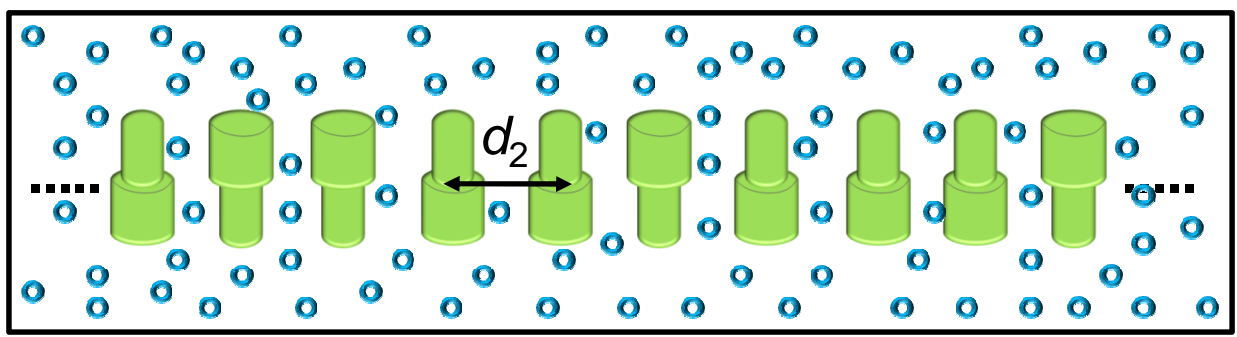

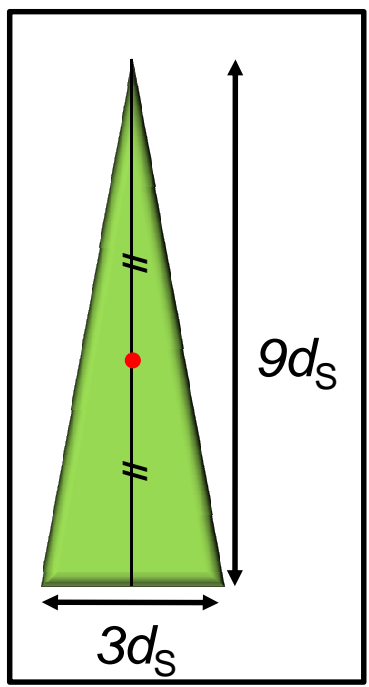

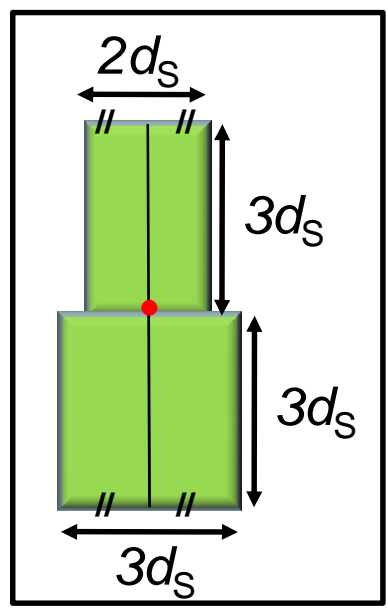

FIG. 2

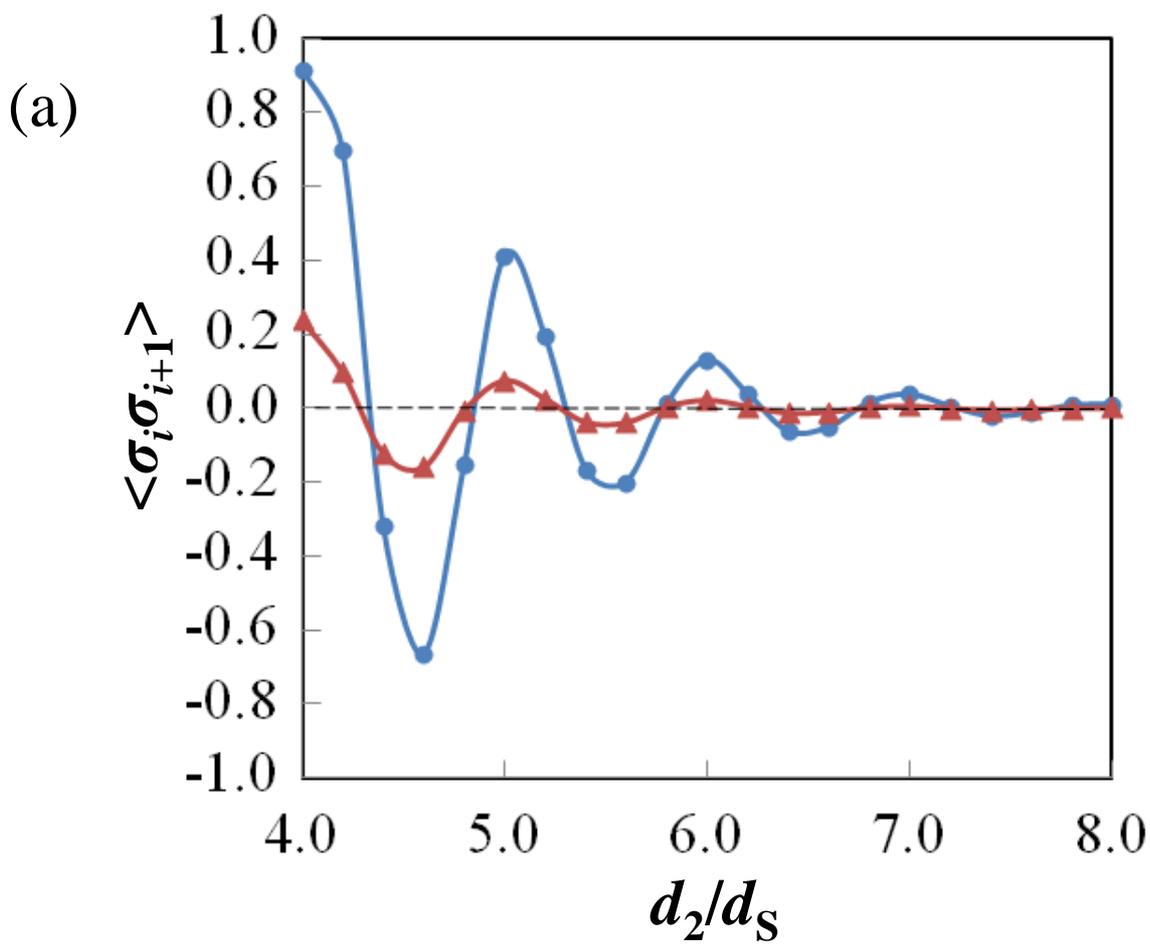
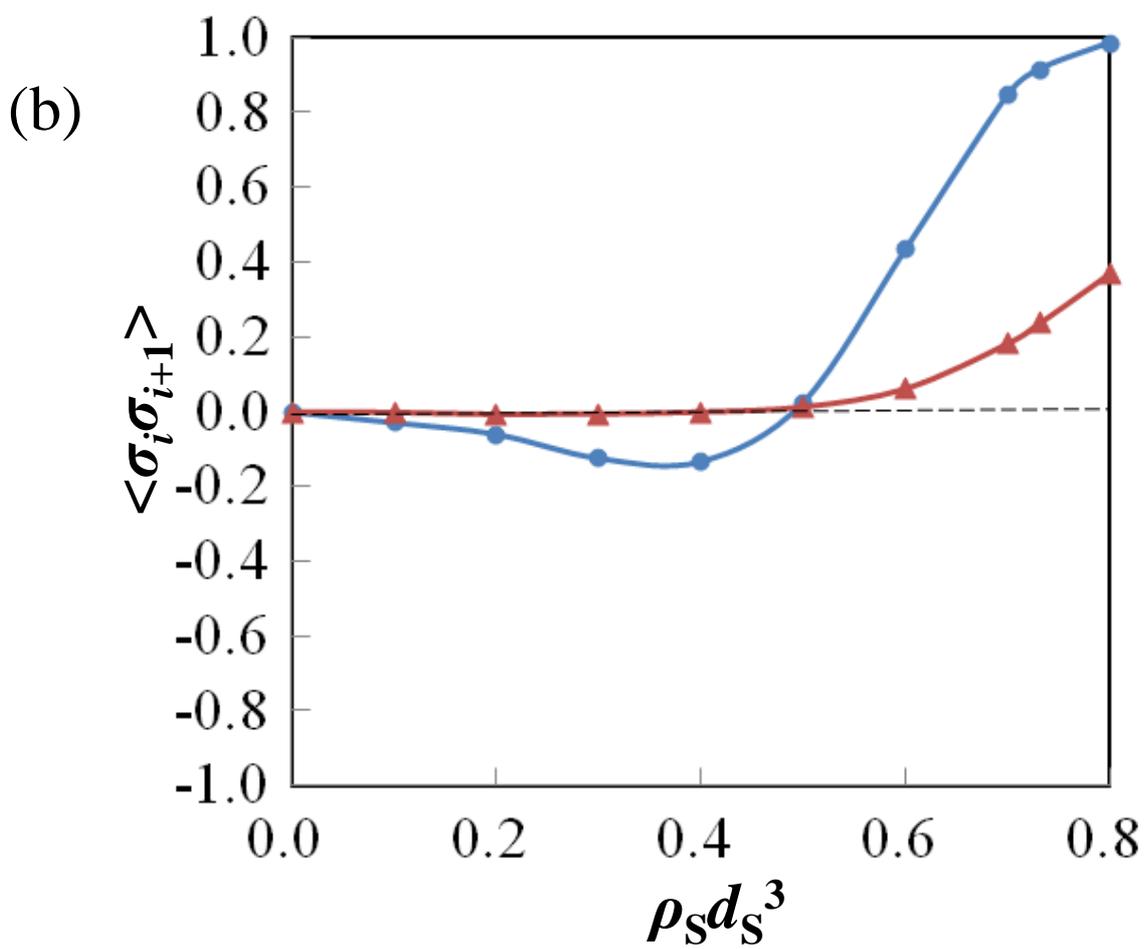

FIG. 3

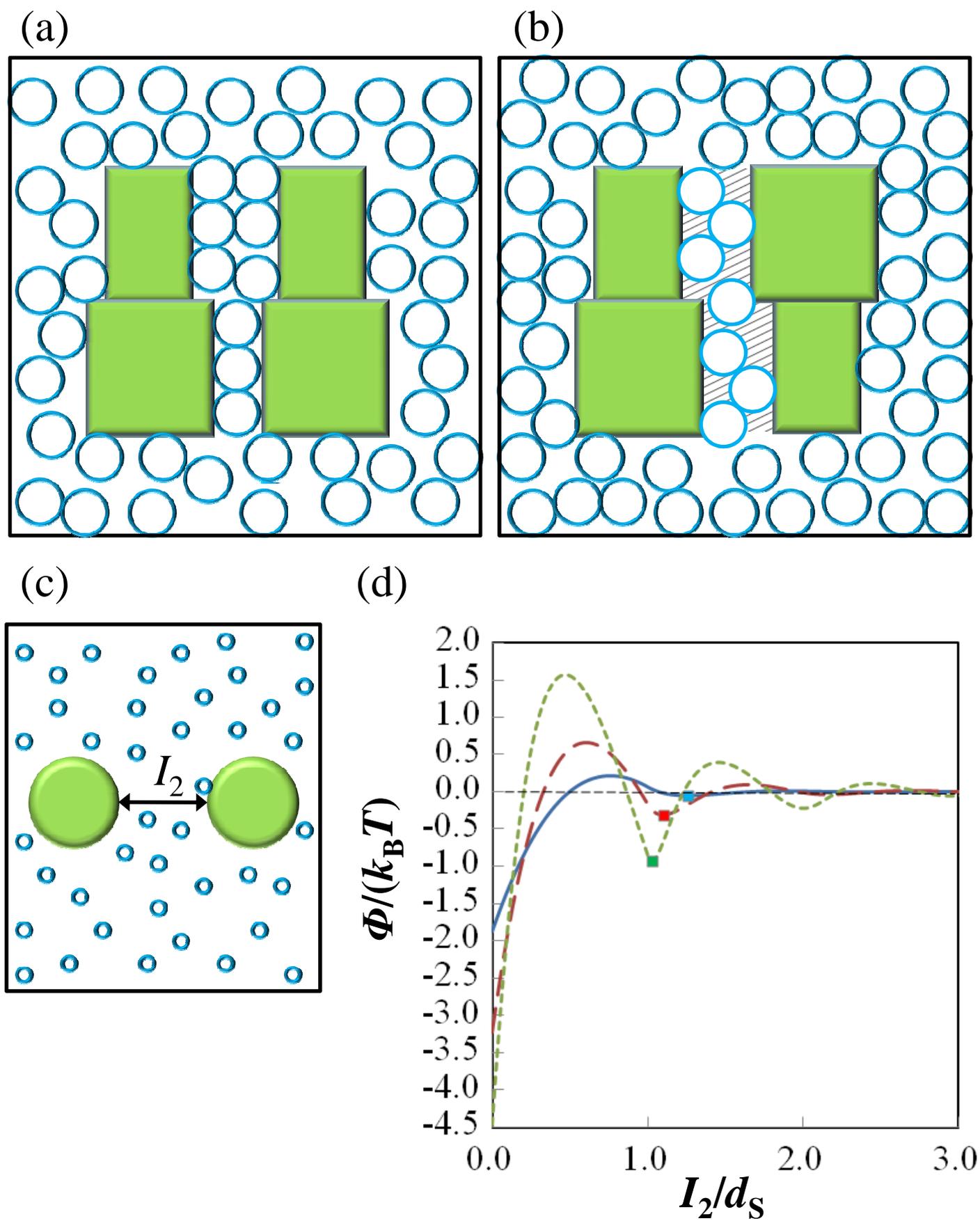

Fig. 4

# Supplemental Material for
# "Theoretical study of solvent-mediated Ising-like system: A study for future nanotechnology"

Ken-ich Amano[1*]

[1]*Graduate School of Energy Science, Kyoto University, Uji, Kyoto 611-0011, Japan*

*Corresponding author: k-amano@iae.kyoto-u.ac.jp.

## 1. Three-dimensional integral equation theory

In the 1D-SI system, the conical and cylindrical solutes are one-dimensionally and regularly placed in the solvent (see Fig. 2), the system of which is all-rigid-body system. Diameter of the solvent sphere ($d_S$) is 0.28 nm, the value of which is that of a water molecule. Bulk density of the solvent is written as $\rho_S$, and $\rho_S d_S^3$ (dimensionless value) is set at 0.7317, the value of which is that for water under the normal condition (1 atm, 298 K).

The two solutes with parallel or antiparallel orientation are regarded as *one* solute here, and it is immersed in the solvent at infinite dilution. The Ornstein-Zernike (OZ) equation in the Fourier space is expressed by [S1-S12]

$$W_{IS}(k_x, k_y, k_z) = \rho_S C_{IS}(k_x, k_y, k_z) H_{SS}(k), \tag{S1}$$

and the hypernetted-chain (HNC) closure equation is written as [S1-S12]

$$c_{IS}(x, y, z) = \exp\{-u_{IS}(x, y, z)/(k_B T)\} \exp\{w_{IS}(x, y, z)\} - w_{IS}(x, y, z) - 1. \tag{S2}$$

Here, subscripts "*I*" and "*S*" represent the *one* solute and the solvent, respectively. *c* is the direct correlation function, *h* the total correlation function, *u* the two-body interaction potential, and $w = h - c$. The capital letters (*C*, *H*, and *W*) represent the Fourier transforms. When there are no overlap and contact between the solute and the solvent particle, $\exp\{-u_{IS}(x, y, z)/(k_B T)\}$ is 1.0 ($k_B$ and *T* are the Boltzmann constant and the absolute temperature, respectively). On grid points where a solvent particle and the solute overlap, it is zero. On those where a solvent particle is in contact with the solute, it is set at 0.5. $H_{SS}(k)$ ($k^2=k_x^2+k_y^2+k_z^2$), which is calculated using the radial symmetric integral equation theory for spherical particles with the HNC approximation [S13], is part of the input data. We emphasize that the OZ equation is *exact*. Only the HNC closure neglecting the bridge function is approximate.

The numerical procedure is summarized as follows:

(1) $u_{IS}(x, y, z)$ is calculated at each 3D grid point.
(2) $w_{IS}(x, y, z)$ is initialized to zero.
(3) $c_{IS}(x, y, z)$ is calculated from Eq. (S2), and $c_{IS}(x, y, z)$ is transformed to $C_{IS}(k_x, k_y, k_z)$ using the 3D fast Fourier transform (3D-FFT).
(4) $W_{IS}(k_x, k_y, k_z)$ is calculated from Eq. (S1), and $W_{IS}(k_x, k_y, k_z)$ is inverted to $w_{IS}(x, y, z)$ using the 3D-FFT.
(5) steps (3) and (4) are repeated until the input and output functions for $w_{IS}(x, y, z)$ become identical within convergence tolerance.

The grid spacing ($\Delta x$, $\Delta y$, and $\Delta z$) is set at $0.1d_S$, and the grid resolution ($N_x \times N_y \times N_z$) is $512 \times 512 \times 512$. It has been verified that the spacing is sufficiently small and the box size ($N_x\Delta x$, $N_y\Delta y$, $N_z\Delta z$) is large enough for the correlation functions at the box surface to be essentially zero. Once the correlation functions ($h_{IS}$ and $c_{IS}$) are calculated, solvation free energy of the *one* solute ($\mu_2$) can be obtained from the 3D Singer-Chandler formula [S14],

$$\mu_2/(k_B T) = \rho_S \iiint \{h_{IS}(x,y,z)^2/2 - c_{IS}(x,y,z) - h_{IS}(x,y,z)c_{IS}(x,y,z)/2\} dxdydz . \tag{S3}$$

The integral equation theory with the HNC approximation was shown to give quite satisfactory results even in a quantitative sense for the following examples: potential of mean force (PMF) between large spherical solutes immersed in pure solvent calculated by the radial symmetric integral equation theory [S13]; PMF between large spherical and nonspherical solutes immersed in pure solvent calculated by the 3D integral equation theory [S4]; and PMF between large spherical solutes immersed in a multicomponent solvent calculated by radial symmetric integral equation theory [S15]. The reliability test was performed by comparing results obtained by the integral equation theory with the HNC approximation with those obtained by the density functional theory [S7,S15], which was shown to give the results indistinguishable from those by an MD simulation for rigid-body systems, or from the integral equation theory with the exact bridge functions [S13].

## 2. Discussion of an approximation

We considered only interactions between the nearest-neighbor solutes to solve the partition function of the 1D-SI system. That is, interactions between the $i$-th and $(i+n)$-th ($n \geq 2$) solutes are ignored. According to the approximation (see Eqs. (5) and (6)), solvation free energy is changed by $\Delta\mu_{PA}/2$ from the standard value of $\mu_{PS}+N(\mu_{2P}+\mu_{2A})/2$, when the two solutes are parallel. In contrast, when the two solutes are antiparallel, the change of the solvation free energy is $-\Delta\mu_{PA}/2$. Therefore, when the solute is sandwiched between the closest two solutes and the three spins are parallel (↑↑↑), the solvation free energy change is simply estimated to be $\Delta\mu_{PA}$, because there are two parallel couples in (↑↑↑). On the other hand, when the three spins are antiparallel (↑↓↑), the solvation free energy change is $-\Delta\mu_{PA}$ due to the two antiparallel couples in (↑↓↑). That is, effect of the sandwich, which is considered to be unignorable in some cases [S10,S11], is approximated to be nothing here. In this section, we check validity of the approximation.

To check the validity, we straightforwardly immerse the three (conical or cylindrical) solutes in the solvent and calculate solvation free energy of the three solutes ($\mu_3$). The calculation is performed by regarding the three solutes as *one* solute [S5]. We test two conditions, (↑↑↑) and (↑↓↑), and $\mu_3$ calculated by the approximation method and that by the straightforward method are compared. The comparison is conducted at arbitrary $d_2$ ($d_2$=4.0, 4.4, 4.6, 5.0, 5.4, 5.6, 6.0, 6.4, 6.6, 7.0, 7.4, 7.6, and 8.0), where $\rho_S d_S^3$ is set at 0.7317. The results of the conical and cylindrical solutes are shown in Figs. S1(a) and S1(b), respectively. As shown in Fig. S1, the two methods produce quantitatively (almost) the same values, differences of which are all negligibly small. That is, the sandwich effect can be ignored and the approximation is proven to be valid.

## References
[S1] D. Beglov and B. Roux, J. Chem. Phys. **103**, 360 (1995).

[S2] M. Ikeguchi and J. Doi, J. Chem. Phys. **103**, 5011 (1995).

[S3] M. Kinoshita and T. Oguni, Chem. Phys. Lett. **351**, 79 (2002).

[S4] M. Kinoshita, J. Chem. Phys. **116**, 3493 (2002).

[S5] M. Kinoshita, Chem. Phys. Lett. **387**, 47 (2004).

[S6] M. Kinoshita, Chem. Phys. Lett. **387**, 54 (2004).

[S7] Y. Harano and M. Kinoshita, Chem. Phys. Lett. **399**, 342 (2004).

[S8] Y. Harano and M. Kinoshita, Biophys. J. **89**, 2701 (2005).

[S9] M. Kinoshita, Chem. Eng. Sci. **61**, 2150 (2006).

[S10] K. Amano, T. Yoshidome, M. Iwaki, M. Suzuki, and M. Kinoshita, J. Chem. Phys. **133**, 045103 (2010).

[S11] K. Amano and M. Kinoshita, Chem. Phys. Lett. **488**, 1 (2010).

[S12] K. Amano and M. Kinoshita, Chem. Phys. Lett. **504**, 221 (2011).

[S13] M. Kinoshita, S. Iba, K. Kuwamoto, and M. Harada, J. Chem. Phys. **105**, 7177 (1996).

[S14] S. J. Singer and D. Chandler, Molec. Phys. **55**, 621 (1985).

[S15] R. Roth and M. Kinoshita, J. Chem. Phys. **125**, 084910 (2006).

**Figure Caption**

FIG S1. Comparison between the approximation and straightforward methods. (a) and (b), solvation free energies of the three conical and cylindrical solutes, respectively. They are divided by $k_\mathrm{B}T$, and calculations are performed under the condition $\rho_\mathrm{S} d_\mathrm{S}^3 = 0.7317$. Blue and red circles represent that of (↑↑↑), calculations of which are performed by the approximation and straightforward methods, respectively. Green and orange circles represent that of (↑↓↑), which are calculated by the approximation and straightforward methods, respectively.

(a)

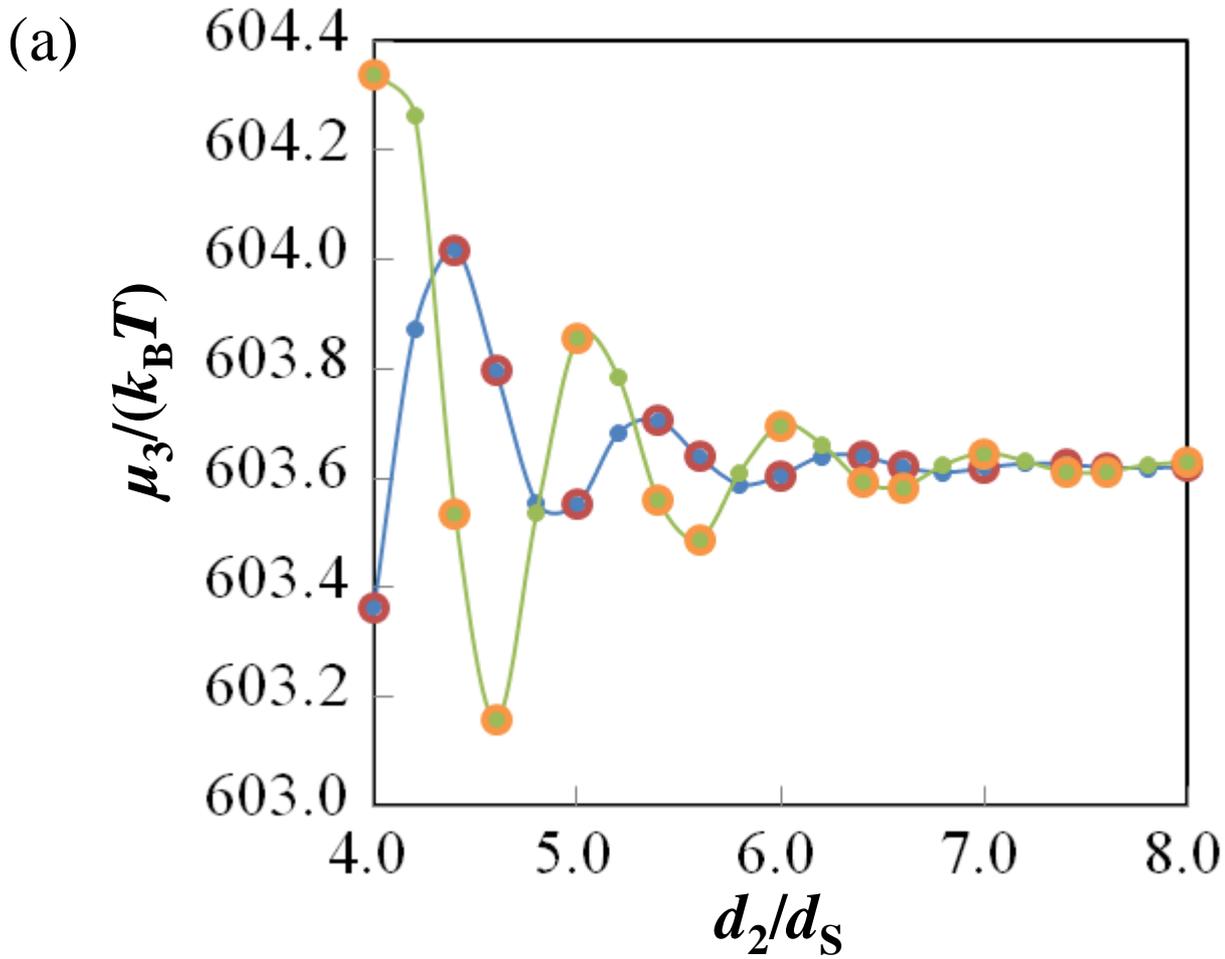

(b)

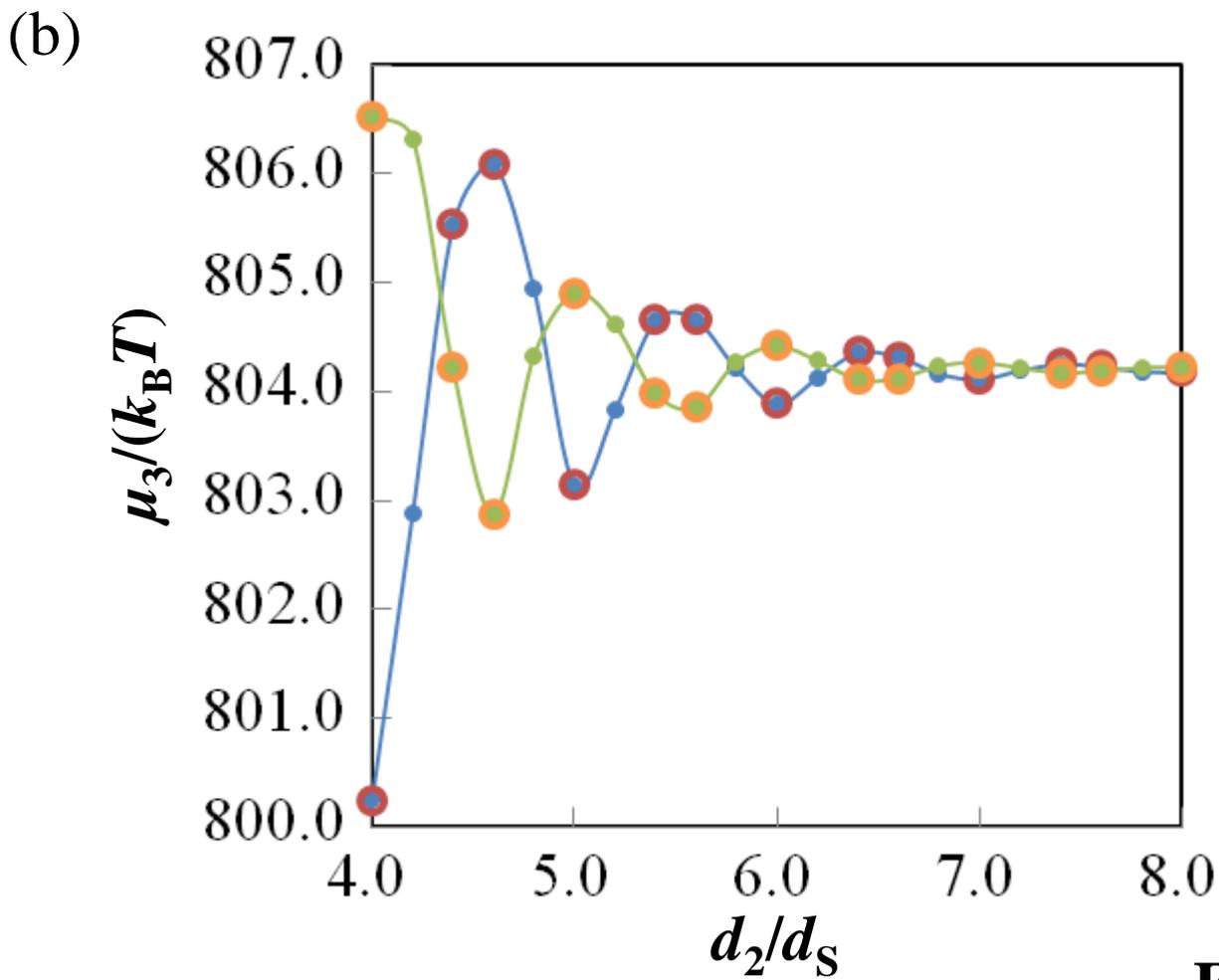

**FIG. S1**